\documentclass[twocolumn,amsmath,amssymb,prl,aps]{revtex4-1}
\usepackage{graphicx}
\usepackage{amsmath,amssymb}
\usepackage{mathrsfs}
\usepackage{color}
\usepackage{afterpage}
\usepackage[version=3]{mhchem}
\usepackage{natbib}
\usepackage{soul}
\usepackage[caption=false]{subfig}
\usepackage{array}
\usepackage{multirow}
\usepackage[colorlinks, citecolor=blue,urlcolor=blue, linkcolor=blue, bookmarks=false]{hyperref}
\hypersetup{colorlinks=true , citecolor=blue, urlcolor=blue, linkcolor=blue}
\begin{document}

\title{Probing the uniaxial strain-dependent valley drift and Berry curvature in monolayer MoSi$_2$N$_4$ }

\author{Sajjan Sheoran\footnote{phz198687@physics.iitd.ac.in} and Saswata Bhattacharya\footnote{saswata@physics.iitd.ac.in}} 
\affiliation{Department of Physics, Indian Institute of Technology Delhi, New Delhi 110016, India}
\date{\today}

\begin{abstract}
   We use \textit{ab initio} calculations and theoretical analysis to investigate the influence of in-plane strain field on valley drifts and Berry curvatures in the monolayer MoSi$_2$N$_4$, a prototypical septuple atomic layered two-dimensional material. The low energy electron and hole valleys drift far off the K/K$'$ point under uniaxial strains. The direction and strength of valley drift strongly depend on the nature of the charge carrier and uniaxial strain with a more substantial response along the zigzag path. Our findings are governed by the interplay between microscopic orbital contribution and symmetry lowering. The changing geometric properties of Bloch states affect the Berry curvatures and circular dichroism. Specifically, Berry curvature dipole is significantly enhanced under the tensile strain along armchair and zigzag directions. Meanwhile, the particle-hole asymmetry arising from non-equivalent electron and hole valley drifts relax the selection rules, thus reducing the degree of circular polarization up to $\sim$0.98. Therefore, strain engineering of valley physics in the monolayer MoSi$_2$N$_4$ is of prime importance for valleytronics.
\end{abstract}
%\pacs{}
%\keywords{DFT, valley drift, Berry curvature, circular dichroism}
\maketitle

\textit{Introduction:-} %Valleytronics has provoked great attention in recent years, where valley degree of freedom are exploited for information encoding and processing. Valley refers to presence of multiple energy extremal point in the the Brillouin zone (BZ).
 Two-dimensional (2D) layered semiconductors are essential in manipulating the valley degrees of freedom for potential valleytronics devices~\cite{schaibley2016valleytronics,vitale2018valleytronics}. Several intriguing phenomena are experimentally accomplished, such as the valley Hall effect~\cite{gorbachev2014detecting, mak2014valley,lee2016electrical,PhysRevLett.108.196802}, valley excitons~\cite{chernikov2014exciton,ye2014probing}, valley Zeeman and AC Stark effects~\cite{rostami2015valley}. MoSi$_2$N$_4$ is a newly discovered 2D material that has attracted significant scientific attention~\cite{hong2020chemical, nsr}. %Monolayer MoSi$_2$N$_4$ have excellent ambient stability (no structural change under ambient conditions even after 6 months), large synthesis size (upto 15mm$\times$15mm)~\cite{hong2020chemical}. 
 The monolayer MoSi$_2$N$_4$ has pair of Dirac valleys at corners of the Brillouin zone (BZ) connected by the time-reversal symmetry operation and constitutes a binary index for low energy carriers~\cite{wang2021intercalated,yang2021valley,li2020valley,wang2021electronic,islam2021tunable,ai2021theoretical,yin2022emerging}. The breaking of inversion symmetry and high spin-orbit coupling leads to valley contrasting features in the spin splitting, berry curvatures, and optical circular dichroism~\cite{li2020valley,sheoran2023coupled}. Thus, valley polarization can be controlled by optical, transport, and magnetic interactions. %In addition, the valleys in the monolayer MoSi2As4 are multiple-folded, enabling multiple information processing~\cite{yang2021valley}. 
 Theoretical calculations predict the high electron/hole mobility up to ~270/1200 cm$^2$V$^{-1}$s$^{-1}$ for monolayer MoSi$_2$N$_4$, which is nearly six times larger than that of monolayer MoS$_2$ predicted~\cite{hong2020chemical}. The excellent stability, carrier mobility, suitable band gap, and protection of capped SiN layer from environmental disturbance promise an advantage over MoS$_2$ in valley transport properties, especially valley filtering, valley Hall effect, and valley coupled spin Hall effect~\cite{hong2020chemical,wu2022prediction}.

Strain engineering is used successfully to improve the performance of monolayer devices by modifying the band dispersion. Some groundbreaking discoveries related to strains include band gap reduction~\cite{zhu2013strain,castellanos2013local,rostami2015theory,rostami2016edge}, direct to indirect band transition~\cite{conley2013bandgap}, funneling of photogenerated excitons~\cite{san2016inverse}, tunnel resistance modulation~\cite{fu2013mechanically}, and enhancement in Rashba splitting~\cite{gentile2015edge,PhysRevMaterials.6.094602,D1MA00912E,bhumla2021origin}. The uniaxial strain can also lower the symmetry, inducing the band splitting and modulation of Berry curvature dipole and valley magnetization~\cite{son2019strain}. The strain superlattices can open the significant energy gap and shift the Dirac points in graphene~\cite{guinea2010energy}. Furthermore, several experimental and theoretical investigations predict the direct to indirect band gap transition in MoS$_2$ under the tensile strain between 2 to 3\%~\cite{zhu2013strain,shin2016indirect, jena2019valley}. Most importantly, 2D materials are able to sustain significant reversible elastic strain, i.e., graphene (25\%) ~\cite{lee2008measurement,kim2009large,cao2020elastic} and MoS$_2$ (11\%)~\cite{bertolazzi2011stretching,liu2022approaching,he2013experimental}. % The strong in-plane ionic covalent bonding in monolayer  arising from the orbital overlap between Mo-$d$ and S-$p$ orbitals sustains the strain up to 11\% with breaking strength of 23 GPa.
The strain engineering of monolayer MoSi$_2$N$_4$ has started to gain attention recently~\cite{li2020valley,liang2022highly,mortazavi2021exceptional,kang2021second}. Thus, the mechanical tunability of its valley properties provides an ideal avenue to prospect.

This study demonstrates strain-induced electron and hole valley drifts and their effects in monolayer MoSi$_2$N$_4$. We theoretically apply in-plane lattice deformation up to 10\% along three crystallographic non-equivalent directions, i.e., armchair (AC), intermediate (IM), and zigzag (ZZ). The energy valley drifts are responsive to the strain directions. For example, drifts are more pronounced for strain along the ZZ direction than the IM and AC directions. Additionally, electron valleys are more sensitive to the deformation when compared to the hole valleys. The group theoretical perspective and microscopic orbital contribution explain the underlying mechanism behind the asymmetric nature of valley drifts. Afterward, we provide the effect of asymmetric valley drifts on Berry curvature distribution and optical circular dichroism. 

\begin{figure}[htp]
	\includegraphics[width=7.5cm]{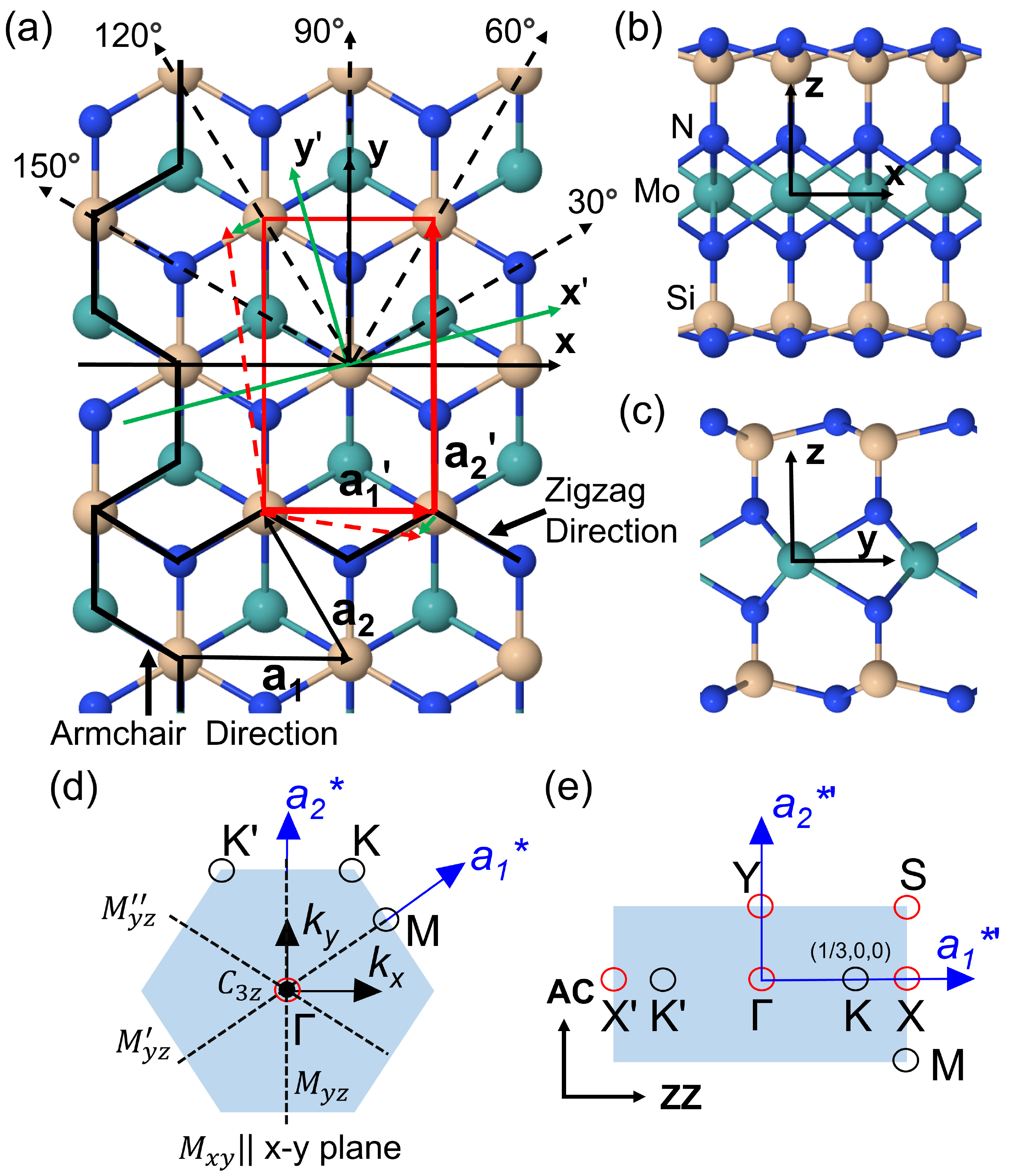}
	\caption{(a) Top and [(b),(c)] side views of monolayer MoSi$_2$N$_4$. The x, y, and z axes are shown in each case. $a_1$ and $a_2$ are the primitive lattice vectors, while $a_1'$ and $a_2'$  are lattice vectors of the orthorhombic cell. The rotated Cartesian frame for strain along an arbitrary direction is shown using green lines. After applying strain along $x'$ direction and allowing relaxation along perpendicular $y'$ direction, the orthorhombic unit cell is shown using red dashed lines. The effect of strain on the strained lattice is exaggerated for clear illustration and is much smaller for our considered range. x and y directions are zigzag (ZZ) and armchair (AC), respectively. The atomic arrangement is ZZ along 0$^\circ$, 60$^\circ$ and 120$^\circ$, whereas AC along 30$^\circ$, 90$^\circ$ and 150$^\circ$  directions. The first (d) hexagonal and (e) orthorhombic BZ are shown. $M_{xy}$ and $M_{yz}$ are the mirror planes, and $C_{3z}$ is the three-fold rotation axis. The high symmetry points $\Gamma$, M, and K of hexagonal BZ are mapped in the orthorhombic BZ (see Eq.~\ref{e1}).}
	\label{p1}
\end{figure}

\textit{Calculation details and methods:-} Density functional theory~\cite{hohenberg1964inhomogeneous,kohn1965self} calculations were performed using the projector augmented wave method as implemented in the Vienna ab initio simulation package (VASP)~\cite{kresse1996efficiency,kresse1996efficient}. The Perdew-Burke-Ernzerhof (PBE) pseudopotentials were used for exchange-correlation potential~\cite{perdew1996generalized}. The BZ was sampled using the Monkhorst-Pack method with a spacing of 0.02 \AA$^{-1}$~\cite{monkhorst1976special}. A vacuum of 20 {\AA} along the $z$-direction was used  to avoid artificial interaction between periodic images. The Wannier representations were obtained by projecting the Bloch states from first principle calculations on Mo-$3d$, N-$2s$, N-$2p$ and Si-$3p$ orbitals~\cite{mostofi2008wannier90}.

The monolayer MoSi$_2$N$_4$ has the hexagonal lattice structure built by septuple atomic layers in the sequence N-Si-N-Mo-N-Si-N (see Fig.~\ref{p1}). The lattice can be regarded as 2H-MoN$_2$ monolayer sandwiched between two silicene-like Si-N monolayers~\cite{hong2020chemical}. There are two alternative choices for the unit cell of monolayer MoSi$_2$N$_4$ having the P$\overline{6}m2$ space group. As shown in Fig. 1(a), the choices are the P$\overline{6}m2$:$h$ (7 atoms hexagonal lattice) and P$\overline{6}m2$:$o$ (14 atoms orthorhombic lattice). It is very efficient to perform DFT calculations in the P$\overline{6}m2$:$h$ setting. However, the P$\overline{6}m2$:$o$ is of great convenience in terms of symmetry perspective for strained lattice along AC or ZZ directions, and therefore considered in our study. Firstly, we have mapped the arbitrary k-point ($k_1$, $k_2$) of hexagonal BZ on the basis of orthorhombic reciprocal vectors. The results are expressed as (see section I of supplemental material (SM)~\cite{SuMa} for more details)
 \begin{equation}
 	(k_1, k_2)_h \implies (k_1, k_1+2k_2)_o
 	\label{e1}
 \end{equation}
The location of $\Gamma$, M, and K points are shown in Fig.~\ref{p1}(d), and corresponds to (0, 0, 0), (1/2, 1/2, 0), and (1/3, 0, 0) in orthorhombic BZ, respectively. Firstly, we have fully relaxed the structure using the conjugate gradient algorithm until the force on every atom is smaller than 1 meV/{\AA}. The simulated lattice parameter ($a_1$) was found to be 2.912 {\AA}, which is in good agreement with the previous experimental~\cite{hong2020chemical,nsr} and theoretical studies~\cite{wang2021intercalated,yang2021valley,li2020valley,wang2021electronic,islam2021tunable,ai2021theoretical}.

To simulate the strained lattice of monolayer MoSi$_2$N$_4$ along $x$ (ZZ) and $y$ (AC) directions, we change the corresponding lattice vector to satisfy the relation, $\epsilon_x=\frac{a_1''-a_1'}{a_1'}$ and $\epsilon_y=\frac{a_2''-a_2'}{a_2'}$, where $a_1'(a_2')$ and $a_1''(a_2'')$  are the lattice constant along $x(y)$ direction without and with strain respectively (see Fig~\ref{p1}(a)). For the application of strain along an arbitrary direction, we rotate the coordinate system such that $x'$ points along the direction of strain. After that, we apply strain, relax the structure along $y'$ direction, and then return to the original Cartesian system. Due to the $C_{3z}$ and $M_{yz}$ symmetry, only the direction between $\theta=0^{\circ}$ and $30^{\circ}$ are of interest, where $\theta$ is measured anticlockwise with respect to the ZZ direction. Therefore, we have considered $\theta=0^{\circ}$ (ZZ), $\theta=15^{\circ}$ (IM), and $\theta=90^{\circ}$ ($\equiv30^{\circ}$, AC) directions. For the uniaxially strained lattice, $D_{3h}$ symmetry reduces to $C_{2h}$, containing a two-fold rotation operation ($C_{2z}$) and mirror symmetry containing the $x$-$y$ plane. Moreover, additional diagonal mirror planes arise for the strain along AC and ZZ directions, leading to $D_{2d}$ point group symmetry.
 
 The elastic energy density for orthorhombic lattices is expressed as~\cite{cadelano2010elastic,cadelano2012effect}
 \begin{equation}
 	U_{ortho}=\frac{1}{2}C_{11}\epsilon_{xx}^2+\frac{1}{2}C_{22}\epsilon_{yy}^2+C_{12}\epsilon_{xx}\epsilon_{yy}+C_{44}\epsilon_{xy}^2.
 	\label{e2}
 \end{equation}
 $C_{ij}$ and $\epsilon_{ij}$ are components of elastic constants and strain tensors, respectively.
 Employing Eq.~\ref{e2}, the direction-dependent Young's modulus and Poisson's ratio are obtained as~\cite{cadelano2010elastic}
  \begin{align}
 	\begin{aligned}
 		Y_{2D}(\theta) &= \frac{\Delta}{C_{11}s^4+C_{22}c^4+(\frac{\Delta}{C_{44}-2C_{12}})c^2s^2}\\
 		\nu_{2D} (\theta) &= \frac{(C_{11}+C_{22}-\frac{\Delta}{C_{44}^2})c^2s^2-C_{12}(s^4+c^4)}{C_{11}s^4+C_{22}c^4+(\frac{\Delta}{C_{44}-2C_{12}})c^2s^2}
 	\end{aligned}
 \end{align}
where $\Delta=C_{11}C_{22}-C_{12}^2$, $c=cos(\theta)$ and $s=sin(\theta)$.
\begin{figure}[htp]
	\includegraphics[width=7.5cm]{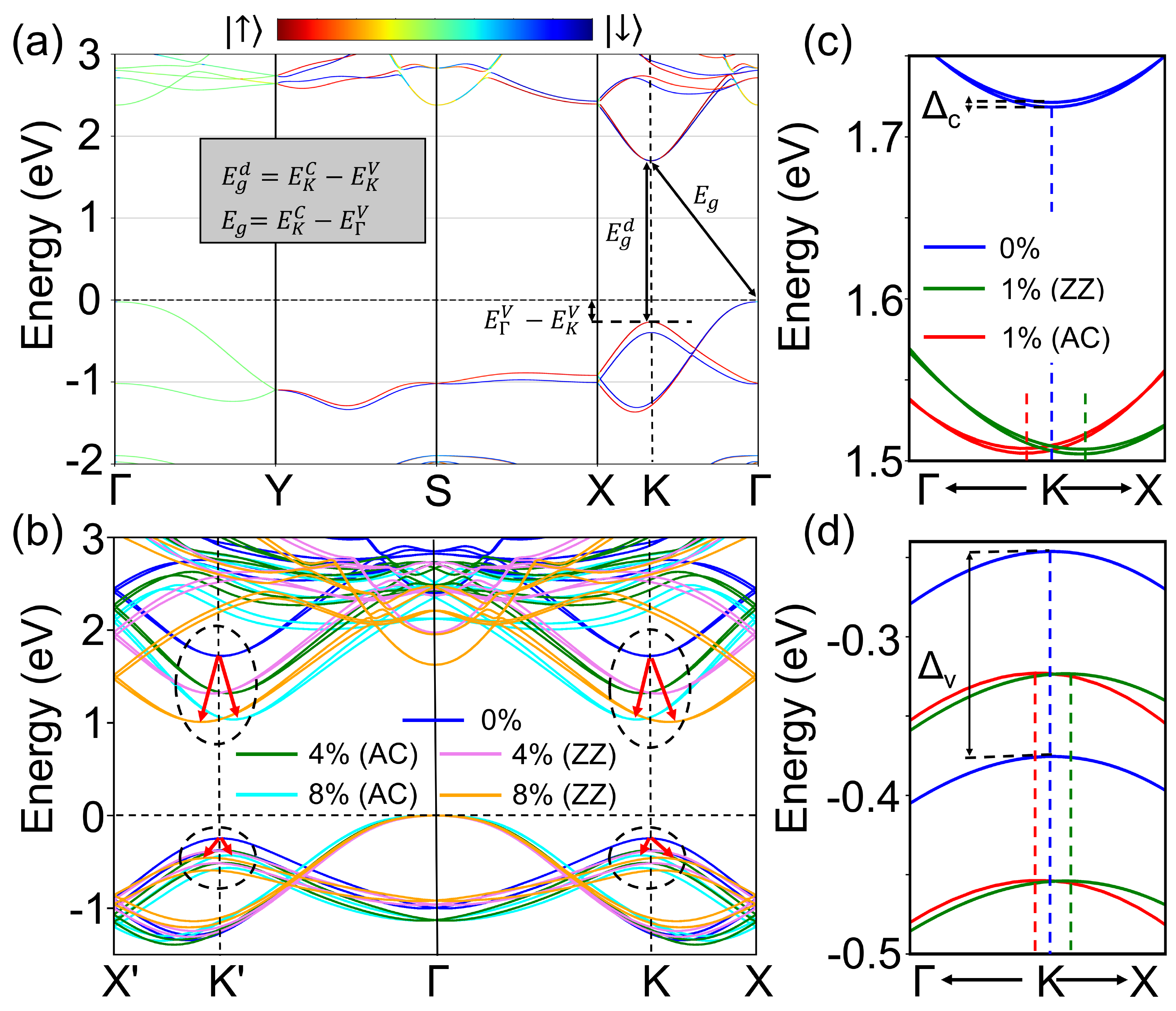}
	\caption{(a) Spin-projected band structures of the monolayer MoSi$_2$N$_4$ in the presence of SOC. The red and blue denote the spin-up and spin-down states, respectively. (b) Evolution of band structures under the uniaxial strain of strength 4\% and 8\% along the ZZ and AC directions. The red arrows represent the valley drifts. (c) and (d) are the zoomed versions of low conduction bands and top valence bands of monolayer MoSi$_2$N$_4$ with 1\% uniaxial stain, respectively. The vertical dashes represent the location of energy extreme in $k$-space}
	\label{p2}
\end{figure}

\textit{Results:-} Spin-projected band structure of monolayer MoSi$_2$N$_4$ is shown in Fig.~\ref{p2}. It has a  band gap of 1.72 eV, having valence band maximum (VBM) and  conduction band minimum (CBM) at the $\Gamma$ and K point, respectively. The calculated band gap is smaller than the experimental value of 1.94 eV~\cite{hong2020chemical} due to self-interaction error, while it is in good agreement with previous theoretical PBE results~\cite{wang2021intercalated,yang2021valley,li2020valley,wang2021electronic}. In Refs~\cite{wu2022prediction,liang2022highly,yadav2022strongly}, band gaps are improved using hybrid HSE06 and many-body perturbation theories (GW, BSE). The results in this work are independent of the choice of functional. Thus, our results are based on the PBE functional. The valence band (VB) at K points lies lower in energy than VBM with an offset ($\Delta_{\Gamma-K}$=$E^V_{\Gamma}$-$E^V_K$) value of 0.24 eV (see Fig.~\ref{p2}(a)). Spins eigenstates are polarized along the $z$-direction except for the $\Gamma$-Y line and can be explained by the symmetry aspects~\cite{wang2021electronic}. Horizontal-mirror symmetry ($M_{xy}$) commutes with the Hamiltonian, therefore eliminating the spin-polarization along the $x$-$y$ plane and preserving along the $z$ direction. For any $k$-point along $\Gamma$-Y direction, $H(k)$ exhibits $C_{2v}$ point group symmetry and does not contain the spin operator leading to degenerate eigenvalues. It has Dirac-type valley bands near the K/K$'$ points. Due to the strong SOC and broken inversion symmetry, valley Fermions exhibit strong spin-valley coupling, valley-selective optical circular polarization, and valley-contrasting Berry curvature~\cite{li2020valley,yang2021valley,wang2021electronic}. The VB valley is the spin split of order $\sim$130 meV, whereas the CB valley is nearly spin degenerate (see section II of SM~\cite{SuMa}). These effects are well explained by the previous DFT calculations~\cite{li2020valley,yang2021valley,wang2021electronic}, tight-binding (TB) models~\cite{wang2021electronic}, and two-band and three-bands $k.p$ models~\cite{li2020valley,yang2021valley}.

 Figure~\ref{p2}(b) shows the evolution of band structures under the strain along AC and ZZ directions. We find a robust strain-valley coupling between the strain and low-energy states. Uniaxial strain shifts the electron and hole valleys away from the K/K$'$ point. Figures~\ref{p2}(c) and \ref{p2}(d) shows the valley drift response of low CB and top VB under the uniaxial strain in the limit of 1\% along ZZ and AC directions. Here, we observe that electron valleys responses are stronger when compared to the hole sector. In addition, the parabolicity of electron bands is more heavily deformed than the hole bands. This phenomenon can be explained by the changing geometric effects and orbital hybridization of relevant Bloch states due to reduced symmetry of the lattice. The strain modifies the scalar potential for the corresponding Bloch bands contributing to the low-energy holes and valley states at the K/K$'$ point.
 \begin{figure*}[htp]
 	\includegraphics[width=12.5cm]{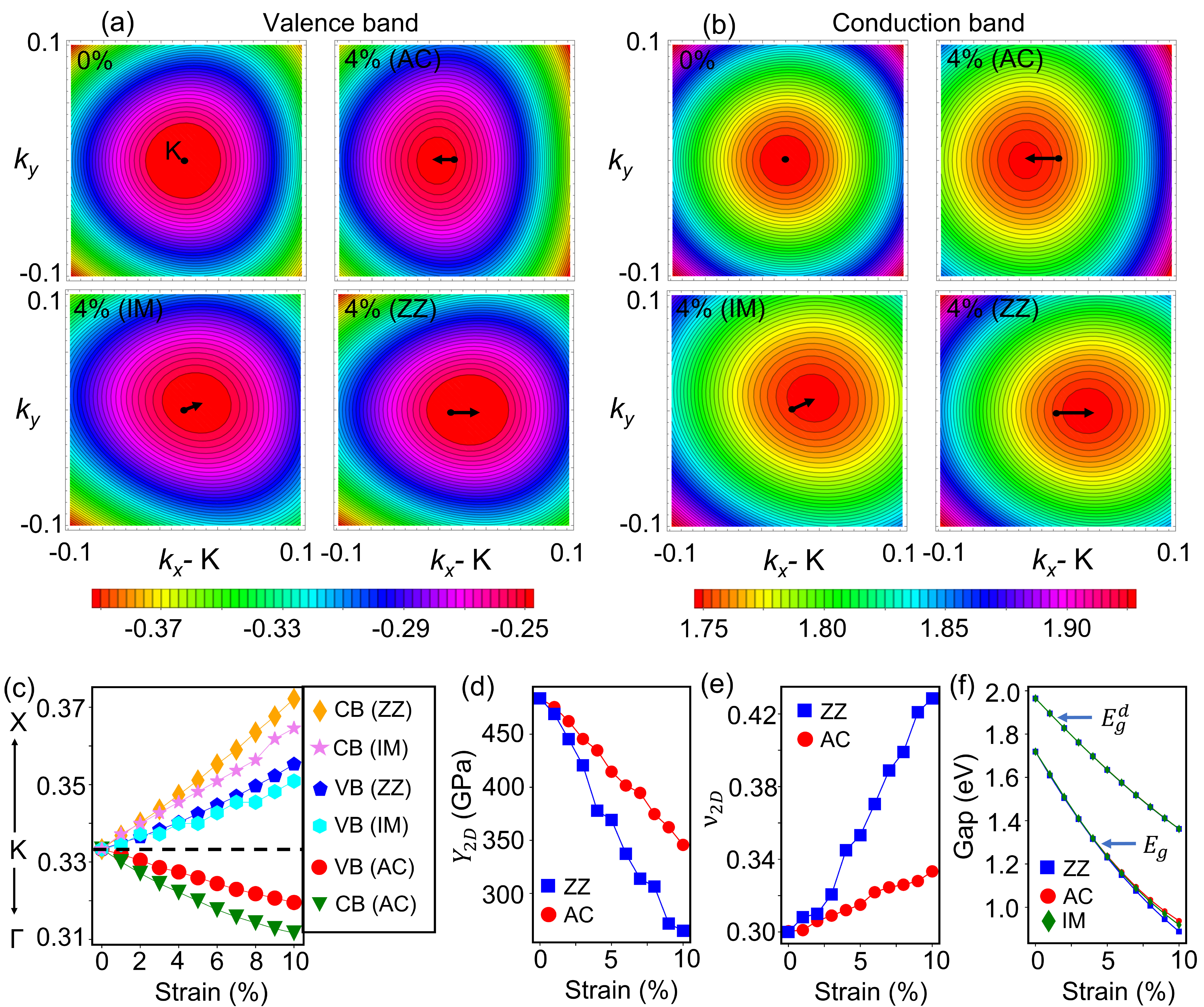}
 	\caption{(a) Contour plot showing isoenergy contours of the valence band around the K point for the unstrained case (0\%) and with the uniaxial strain of 4\% along AC, IM, and ZZ directions of monolayer MoSi$_2$N$_4$. Black arrows show the directions of valley drifts. (b) The same as in (a) for the conduction band. (c) Momentum drift of the CB and VB near the K point. Valley extremum as a function of uniaxial strain applied along the AC, IM, and ZZ direction. Note that only the $k_x$ component of drift along K$\rightarrow$X, cos(30$^\circ$) of the net drift is shown in (c) for the IM case. The $k_y$ component (sin(30$^\circ$) of the net drift) is also there and duly considered in all the calculations. The variation of (d) Young's modulus, (e) Poisson ratio, (f) direct and indirect band gap as a function of strain.}
 	\label{p3}
 \end{figure*}
 
 For a clear illustration of valley drifts, we have plotted the constants energy contours of VB and CB near the K point. The results are shown in Figs.~\ref{p3}(a) and~\ref{p3}(b) (see section III of SM~\cite{SuMa} for energy contours in the entire BZ). Except for the immediate vicinity of the K point, the dispersion is not isotropic. The trigonal warping effect of energy bands is visible, resulting from the $C_{3z}$. The trigonal warping effect is more pronounced for the VB when compared to the CB. When applying the uniaxial strain, an increase in the warping effect near the K point is observed due to the continuous reduction in $C_{3z}$ and translation symmetry along its mutually perpendicular direction. The bands are elliptically warped (due to the appearance of two-fold rotation symmetry) in the immediate proximity of K when uniaxial strain is applied and increases with an increasing strain field. The elliptical warping of band topology is anti-symmetric in such a way that the major axis is along the $k_x$- and $k_y$- direction for strain along ZZ and AC directions, respectively. For lattice strained along the IM direction, the major axis of elliptical warping makes an angle of 30$^\circ$. The anisotropic nature of electronic bands leads to different effective masses, charge transport, and optical absorption in different directions. In addition, the energy extreme of the valley drifts away from the K point. In general, the direction of valley drift is $2\theta$ (with respect to K$\rightarrow$X direction) when strain is applied along $\theta$ direction. Particularly, the direction of valley drift is K$\rightarrow$$\Gamma$ and K$\rightarrow$X for AC and ZZ direction, respectively (shown by black arrows in Fig.~\ref{p3}(a) and~\ref{p3}(b)). To quantify the valley drifts, we plot the valley extreme as a function of strain in Fig.~\ref{p3}(c). The VB drift off the K point with the rate 1.38$\times$10$^{-3}(\frac{2\pi}{a})/\%$ for strain along the AC direction, whereas the rate of drift is 2.19$\times$10$^{-3}(\frac{2\pi}{a})/\%$ for the ZZ direction. For uniaxial strain along the IM direction, drift is in-between that of AC or ZZ direction with the rate of 1.99$\times$10$^{-3}(\frac{2\pi}{a})/\%$. Similarly, the rate of electron valley drifts are  2.17$\times$10$^{-3}(\frac{2\pi}{a})/\%$, 3.43$\times$10$^{-3}(\frac{2\pi}{a})/\%$ and 3.89$\times$10$^{-3}(\frac{2\pi}{a})/\%$ for strain along the AC, IM, and ZZ directions, respectively. The electron valley drifts are nearly $\sim$1.7 times that of hole valley drifts. Similar valley drifts are observed in free-standing graphene($\sim$0.6$\times$10$^{-3}(\frac{2\pi}{a'})/\%$)~\cite{mohr2009two} and monolayer MoS$_2$($\sim$4.5$\times$10$^{-3}(\frac{2\pi}{a''})/\%$)~\cite{zhang2013giant,jena2019valley}. The smaller valley drifts in monolayer MoSi$_2$N$_4$ than MoS$_2$ provide an additional advantage for valley-based applications.
 
 Here, we notice that the drift response is more extensive when the lattice is stretched along the ZZ direction. That is because the stretching of the Mo-N bond is asymmetric when the lattice is stretched along AC or ZZ direction (see Figs.~\ref{p1}(b) and~\ref{p1}(c)). To understand this, we plot Young's modulus ($Y_{2D}(\theta)$) and Poisson's ratio ($\nu_{2D}(\theta)$) as a function of strain in Figs.~\ref{p3}(d) and~\ref{p3}(e). For unstrained monolayer MoSi$_2$N$_4$, Young's modulus  485.7 GPa is in good agreement with the experiment value of 491.4 $\pm$ 139.1 GPa~\cite{hong2020chemical} and theoretical value of 479 GPa~\cite{mortazavi2021exceptional}. The drop in Young's modulus is significant when strain is along ZZ direction. Drastic variation in elastic constants leads to significant cell deformation. Therefore electron and hole valley drifts are more pronounced when uniaxial strain is applied along the ZZ direction. 
 
 Figure~\ref{p3}(f) shows the evolution of direct ($E_g^d$) and indirect ($E_g$) band gaps. The band gap energies are red-shifted under the uniaxial tensile strain. The shift in the band gap is nearly direction independent for the smaller strains in the range of 0-5\%. However, band gaps are red-shifted at a faster rate for the larger uniaxial strains along the ZZ directions and are attributed to the more enormous variation in mechanical constants. The red-shift rates of the indirect and direct band gaps are 0.067 eV/$\%$ and 0.53 eV/$\%$, respectively. These unequal rates lead to the enhancement in the valence band energy offset ($\Delta_{\Gamma-K}$). $\Delta_{\Gamma-K}$ increases from 0.24 to 0.43 eV under the uniaxial strain of 10\%. The optical absorption and photoluminescence experiments study these variations in the band gaps and are widely explored for graphene~\cite{lui2010ultrafast,guinea2010energy,kang2021pseudo} and TMDs~\cite{zhu2013strain,mak2010atomically}. 
 
\begin{figure}[htp]
 	\includegraphics[width=7cm]{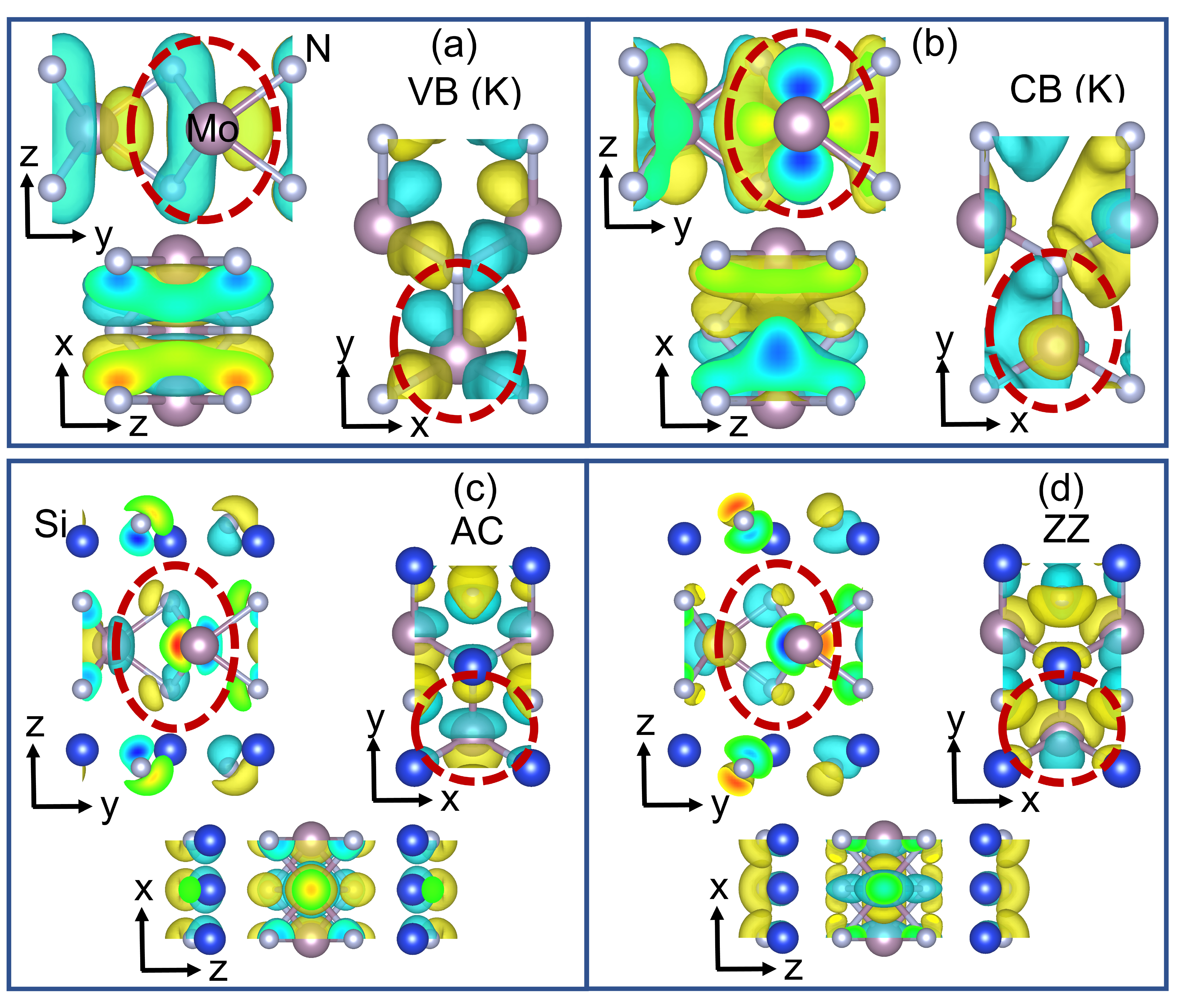}
 	\caption{The real part of the wavefunction of (a) VB and (b) CB at the K point for monolayer MoSi$_2$N$_4$ without strain. Note that in (a) and (b), we show only the Mo-N$_2$ network since the contribution coming from the upper and lower Si-N layer is almost zero. The charge density difference between the unstrained and 4\% strained lattice along the (c) AC and (d) ZZ directions. The electron accumulation is shown in cyan, and electron depletion is shown in yellow.}
 	\label{p4}
 \end{figure}

 The degeneracy of each level is determined by the irreducible representations of point group symmetry of monolayer that is contained in the full rotation symmetry group~\cite{dresselhaus2007group}. 
  \begin{table}
 	\begin{center}
 		\caption{ The irreducible representations and basis functions for the little group $C_{3h} (=C_3\times\sigma_h$) of K/K$'$ point~\cite{dresselhaus2007group}. The sign $\pm$ corresponds to $\pm$K points. The last column contains the energy bands  to which basis functions contribute.}
 		\label{t1}
 		\begin{tabular}{ p{0.8cm}| p{0.1 cm} p{ 0.5cm} p{0.5cm} p{1.2cm} p{3.5cm} p{1cm}}
 			\hline
 			\hline
 			  $C_{3h}$&  & $C_3$ & $\sigma_h$ & $\;\;\;$Mo & $\;\;\;\;\;\;\;\;\;\;\;\;\;\;\;$ N & Band \\ \hline
 			 $A'$ & &1 & 1 &$|\Psi^{\textrm{Mo}}_{2,\mp 2}\rangle$& $\frac{1}{\sqrt{2}}$($|\Psi^{\textrm{N1}}_{1,\mp 1}\rangle + |\Psi^{\textrm{N2}}_{1,\mp 1}\rangle$) & VB  \\
 			 $A''$ & &1 & -1 &$|\Psi^{\textrm{Mo}}_{2,\mp 1}\rangle$& $\frac{1}{\sqrt{2}}$($|\Psi^{\textrm{N1}}_{1,\mp 1}\rangle - |\Psi^{\textrm{N2}}_{1,\mp 1}\rangle$) & CB$+$1  \\
 			 $E'_1$ & &$\omega^{\pm}$ & 1 &$|\Psi^{\textrm{Mo}}_{2,0}\rangle$& $\frac{1}{\sqrt{2}}$($|\Psi^{\textrm{N1}}_{1,\pm 1}\rangle + |\Psi^{\textrm{N2}}_{1,\pm 1}\rangle$) & CB \\
 			 $E'_2$ & &$\omega^{\mp}$ & 1 &$|\Psi^{\textrm{Mo}}_{2,\pm2}\rangle$& $\frac{1}{\sqrt{2}}$($|\Psi^{\textrm{N1}}_{1,0}\rangle - |\Psi^{\textrm{N2}}_{1,0}\rangle$) & CB$+$2 \\
 			 $E''_1$ & &$\omega^{\pm}$ & -1 &$|\Psi^{\textrm{Mo}}_{1,0}\rangle$& $\frac{1}{\sqrt{2}}$($|\Psi^{\textrm{N1}}_{1,\pm 1}\rangle - |\Psi^{\textrm{N2}}_{1,\pm 1}\rangle$) & VB$-$2 \\
 			 $E''_2$ & &$\omega^{\mp}$ & -1 &$|\Psi^{\textrm{Mo}}_{2,\mp 1}\rangle$& $\frac{1}{\sqrt{2}}$($|\Psi^{\textrm{N1}}_{1, 0}\rangle + |\Psi^{\textrm{N2}}_{1, 0}\rangle$) & VB$-$1 \\ \hline

 		\end{tabular}
 	\end{center}
 \end{table} 
 The representation of the a full rotation group will be a reducible representation of the $D_{3h}$ group and can be written as $\Gamma_{l=2}=A'\oplus E' \oplus E''$, where $A'$, $E'$, and $E''$ are the irreducible representations of $D_{3h}$ point group. Therefore, the five-fold degenerate Mo-$d$ orbitals split into three categories at the $\Gamma$ point: $A'(d_{z^2})$, $E'(d_{x^2-y^2},d_{xy})$, and $E''(d_{xz},d_{yz})$ under the effect of trigonal crystal field. The group of wavevector at the K/K$'$ point reduces to the $C_{3h}$. Further, we constructed the Bloch state using the linear combination of atomic orbitals approach, $\Psi^{\eta}_{l,m}(r,k)=\sum_{R_{\eta}}e^{ik.R_{\eta}}Y_l^m(r-R_{\eta})$, where $Y_l^m$ are the spherical harmonics and $R_{\eta}$ is the position of Mo, N and Si atoms. We then identify how the Bloch wave function $\Psi^\eta_{l,m}(r,k)$ transforms under the symmetry operations of $C_{3h}$. Table~\ref{t1} classifies the Bloch state at the BZ corners according to the irreducible representations of $C_{3h}$. The $M_{xy}$ mirror symmetry allows hybridization only between $A'$ and $E'$ orbitals, opening a gap at the K point~\cite{PhysRevLett.108.196802}. As seen from Figs.~\ref{p4}(a) and~\ref{p4}(b), the VB and CB states mainly originate from the hybridization of Mo-$d$ and N-$p$ orbitals. The Bloch state contributing to CB at the K/K$'$ point is $|\Psi^{\textrm{Mo}}_{2,0}\rangle$, whereas the states for VB are $|\Psi^{\textrm{Mo}}_{2,2}\rangle$ and $|\Psi^{\textrm{Mo}}_{2,-2}\rangle$ at the K and K$'$, respectively. The periodic parts of the Bloch wavefunctions for the CB and VB at the K point are predominantly composed of $|d_{z^2}\rangle$ and $|d_{x^2-y^2}$-$id_{xy}\rangle$, respectively. The uniaxial strain modifies the crystal field between metal and Mo-N trigonal coordination environment. Therefore, strain modifies the bandwidths of contributing orbitals and energy of the the Bloch states. The energy of $d$ and $p$ orbitals along the strain direction increases with increasing strain. The out-of-plane orbitals $(d_{z^2})$ are more influenced by the in-plane strain when compared to the in-plane orbital $(d_{x^2-y^2},d_{xy})$. Therefore the effect of strain is more prominent for electron valleys leading to larger valley drift for CB. Stretching of orbitals along the strain direction leads to the charge density redistribution and increases with the increasing strain field. To compare the ZZ and AC directions, we plot the ground state charge density difference between the unstrained lattice and the lattice with 4\% of strain. The charge density difference near Mo and inner N atoms is more prominent in the ZZ case compared to the AC case. Strong charge density redistribution leads to sharper changes in the valley drifts, elastic constants, and band gaps for the lattice stretched along the ZZ direction.
 
\begin{figure}[htp]
	\includegraphics[width=6.5cm]{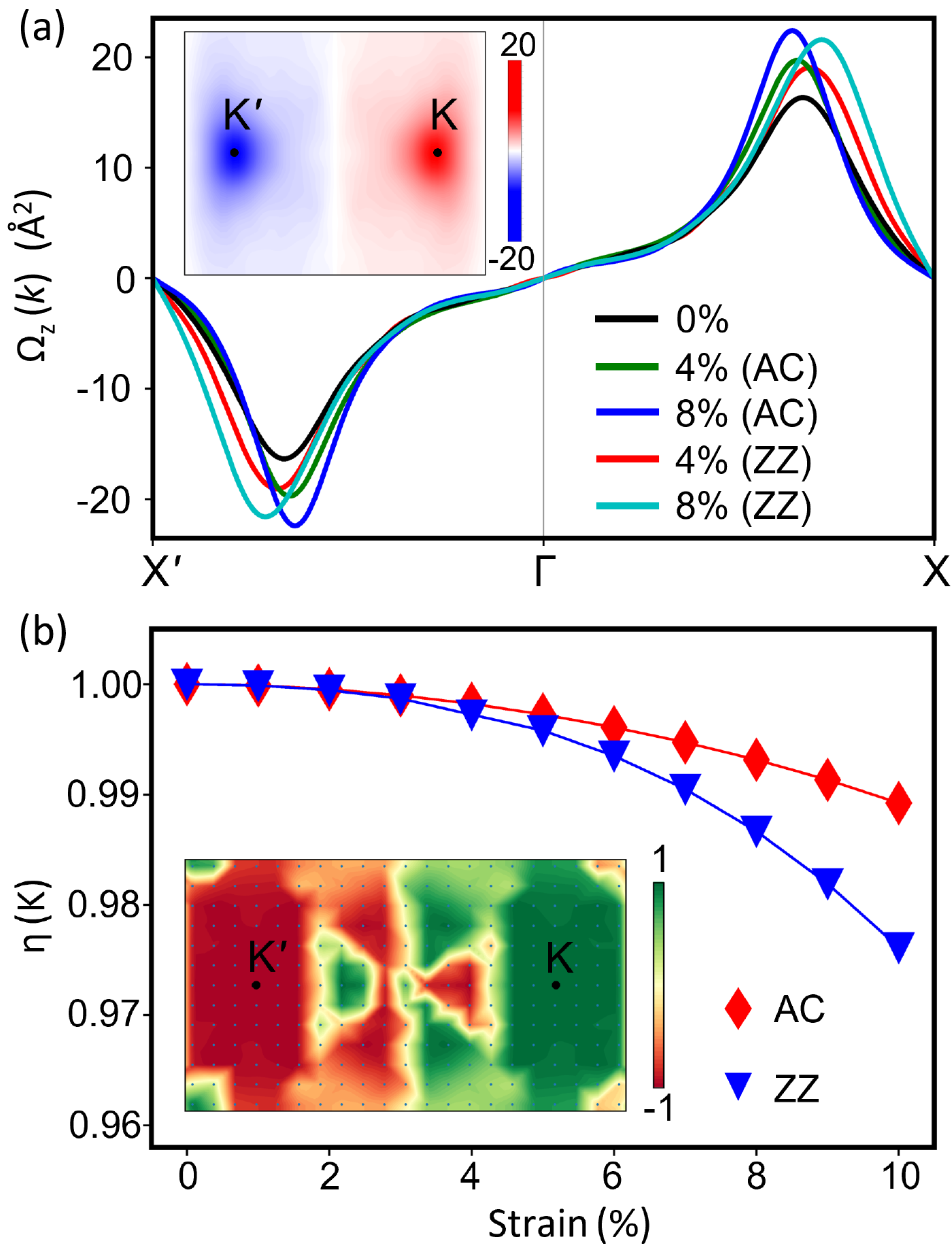}
	\caption{(a) Berry curvature distribution summed for all valence bands along the path X$'$-K$'$-$\Gamma$-K-X in monolayer MoSi$_2$N$_4$. Inset in (a) shows the mapping of Berry curvature under first BZ. (b) The degree of circular polarization of excitation from VB to CB at the K point. Inset in (b) shows the mapping of the degree of circular polarization in the first BZ.}
	\label{p5}
\end{figure}

We further investigate the Berry curvatures in monolayer MoSi$_2$N$_4$ under uniaxial strain. Berry curvature is analogous to the magnetic field in momentum space and has a pivotal impact on the electronic transport properties. For example, the integral of Berry curvature gives Hall conductivity. The Berry curvature is evaluated using the expression
\begin{equation}
	\Omega(k)=i\nabla_{k}\times \langle u(k) | \nabla_{k}| u(k) \rangle
\end{equation}
where $| u(k) \rangle$ is the periodic part of the Bloch wave function. We plot the total Berry curvature of all the occupied bands in Fig.~\ref{p5}(a). Berry curvature distribution is peaked around valleys with opposite values at K and K$'$ points. The contrasting nature of Berry curvatures at the valleys is attributed to the time-reversal symmetry, enabling the separation of charge carriers depending on their valley index. The introduction of strain alters the occupation of different orbitals contributing to Bloch states and leading to modification in the electronic energies and Berry curvature. The Berry curvature profile is strongly modified when the lattice is strained along ZZ or AC direction. The Berry curvature peak drifts away from K/K' point, similar to the energy valley drift. The Berry curvature peak is enhanced up to a factor of $\sim$1.5 under the strain of 8\% along ZZ and AC directions. The increased berry curvature flux density and drift would influence the valley transport when the external transverse electric field is applied since the berry curvature directly enters into the equation of motion and engenders an anomalous velocity term $\sim$$E\times\Omega$. Therein, the strong response of valleys to the strain notably alters the valley-contrasting physics in the monolayer MoSi$_2$N$_4$.

The electron and hole population in two nonequivalent valleys are controlled by circularly polarized optical pumping. The electromagnetic interaction that gives rise to dipole transition is $H_{em}=\frac{e}{mc}P.A$, where $P$ is the momentum operator, and $A$ is the magnetic vector potential of incident light~\cite{dresselhaus2007group}. Then, the probability of dipole transition between the VB ($v(k)$) and CB ($c(k)$) is $|\langle v(k)|H_{em}|c(k)\rangle|^2$. Group theory tells when the dipole transition is allowed or forbidden~\cite{dresselhaus2007group}. The wavefunctions $v(k)$ and $c(k)$ at the K point transforms as the irreducible representations $A'$ and $E'_1$, respectively (see Table~\ref{t1}). $P_{\pm} (=P_x\pm iP_y)$ are the chiral momentum operator representing right/left-hand circularly polarized light. $P_{\pm} (=P_x\pm iP_y)$ transforms as $C_3P_{\pm} C_{3}^{\dagger}=\omega^{\mp}P_{\pm}$~\cite{yang2021valley}. $P_{+}$ transforms according to the irreducible representation $E'_2$. $A'\otimes E'_2 \otimes E'_1$ contains only the fully symmetrical irreducible representation $A'$. Therefore, right-hand circularly polarized light will be absorbed at the K point, whereas absorbance of left-hand circularly polarized light will be forbidden. A similar analysis shows that only left-hand circularly polarized light will be absorbed at the time-reversal conjugate K$'$ point. The degrees of circular polarization ($\eta (k)$) are expressed as
\begin{equation}
	\eta(k)=\frac{|P_{+}|^2-|P_{-}|^2}{|P_{+}|^2+|P_{-}|^2}.
\end{equation}
We calculate the degree of circular polarization using the DFT, and the results agree with the symmetry analysis (see inset in Fig.~\ref{p5}(b)). The transition is exclusively coupled to a specific circular polarization at the valley center for zero strain. After introducing strain, the $\eta(k)$ at the K point decreases due to the continuous reduction of the group of wavevector from $C_{3h}$  to $C_{2h}$. The unequal rates of electron and hole valley drifts soften the valley-selective optical selection rules arising from the $C_3$ symmetry operation due to changes in its optical matrix elements ($\langle v(k)|P_{\pm}|c(k)\rangle$). The $\eta(k)$ reduces up to 0.99 and 0.98 for 10\% uniaxially strained MoSi$_2$N$_4$ along AC and ZZ directions, respectively (see Fig.~\ref{p5}(b)). Therefore, uniaxial strain makes monolayer MoSi$_2$N$_4$ less valley selective, and effects are more prominent along the ZZ direction.

Finally, we emphasize that the analysis presented here can also be manifested in other septuple-layered MA$_2$Z$_4$ (M=Cr, Mo, W, V; A=Si, Ge; Z=N, P, As) materials~\cite{hong2020chemical} and associated Janus structures~\cite{yu2021novel,hussain2022emergence,hussain2023correlation,dey2022intrinsic,guo2022strain,guo2021predicted}. However, the strength of effects can be different. In addition, there are no states other than K/K$'$ near the Fermi level in monolayer MA$_2$P$_4$ and MA$_2$As$_4$~\cite{wang2021intercalated,ai2021theoretical}. Therefore, interference from the different parts of BZ will be minimal, providing an additional advantage over the monolayer MoSi$_2$N$_4$. Furthermore, geometric properties of Bloch states other than Berry curvature can also be modified with the uniaxial strain. For example, orbital magnetic moments ($m_{nk}$) are normal to the plane for 2D materials and have opposite nature for the K and K$'$ valleys. Orbital moments tuning allows control over valley polarization by a vertical magnetic field~\cite{PhysRevB.88.115140}.

\textit{Conclusion:-} In summary, we reported for the first time valley drifts in uniaxially strained monolayer MoSi$_2$N$_4$. By taking strain along AC, IM, and ZZ directions, we confirmed the strain-valley coupling being stronger along the ZZ direction. Additionally, electron valleys show a faster response (nearly 1.7 times) to the strain than hole valleys. These effects noticeably affect the valley's selective optical excitation. We observed a substantial drop in the degree of circular polarization, which can be addressed experimentally by PL spectroscopy. Besides, valley drifts influence the Berry curvature profile around the K/K$'$ point, where an increase in the flux along with drift can induce the valley current in monolayer MoSi$_2$N$_4$. This strong valley asymmetry between valley carriers directly results from changing geometric and orbital effects under directional lattice strain that reduces the lattice symmetry. The theoretical finding in this work may trigger experimental investigations on spin-valley physics in monolayer MoSi$_2$N$_4$ under in-plane lattice strain.

\textit{Acknowledgment:-} S.S. acknowledges CSIR, India, for the senior research fellowship [grant no. 09/086(1432)/2019-EMR-I]. S. B. acknowledges financial support from SERB under a core research grant (grant no. CRG/2019/000647) to set up his High-Performance Computing (HPC) facility ``Veena" at IIT Delhi for computational resources.
\bibliography{ref}

\end{document}

\title{Supplemental Material for\\
	``Probing the uniaxial strain-dependent valley drift and Berry curvature in monolayer MoSi$_2$N$_4$"}
\author{Sajjan Sheoran\footnote{phz198687@physics.iitd.ac.in}, Saswata Bhattacharya\footnote{saswata@physics.iitd.ac.in}} 
\affiliation{Department of Physics, Indian Institute of Technology Delhi, New Delhi 110016, India}
%\date{\today}
\pacs{}
\maketitle
%\section*{1}
\begin{enumerate}[\bf I.]
	
\item C\MakeLowercase{oordinate system}
\item V\MakeLowercase{ariation of spin splitting with uniaxial strain}
\item E\MakeLowercase{nergy contours in the full} BZ

\end{enumerate}

\newpage
\section{C\MakeLowercase{oordinate system}}
We have two choices for the coordinate system of monolayer MoSi$_2$N$_4$ having P$\overline{6}m2$ space group symmetry. The first one is the primitive hexagonal cell (P$\overline{6}m2:h$) having lattice vectors $a_1$ and $a_2$. The angle between $a_1$ and $a_2$ is 120$^\circ$ (see Figure 1 in the main text). Second one is a conventional orthorhombic cell (P$\overline{6}m2:o$) having lattice vectors $a_1'$ and $a_2'$. Note that the conventional cell can also be regarded as a supercell of size   $1\times \sqrt{3}$ supercell. The relation between primitive, and conventional lattice vectors is obtained as
\begin{align}
	\begin{aligned}
		a_1' &= a_1 \\
		a_2' &= a_1+2a_2 \\
		a_3' &= a_3
	\end{aligned}
\end{align}
Suppose that $a_i^*$ and $a_j^{*'}$ are reciprocal lattice vectors corresponding to hexagonal and orthorhombic cells, respectively. According to the definition of reciprocal lattice vectors 
\begin{align}
	\begin{aligned}
		a_i.a_j^* &= 2\pi \delta_{ij} \\
		a_i'.a_j^{*'} &= 2\pi \delta_{ij}
	\end{aligned}
\end{align}
Any general vector in the hexagonal coordinate system ($R_h$) and orthorhombic coordinate system ($R_o$) is expressed as  
\begin{align}
	\begin{aligned}
		R_h &= la_1+ma_2+na_3\\
		R_o &= l'a_1'+m'a_2'+n'a_3'
	\end{aligned}
\end{align}
The general vector is invariant of the choice of coordinate system. Therefore, the relation between $(l,m,n)$ and $(l',m',n')$ is obtained from
\begin{equation}
	R_h=R_o \implies (l'+m')a_1+2m'a_2+n'a_3 = la_1+ma_2+na_3
\end{equation}
Similarly, the general vector in reciprocal space is expressed as
\begin{align}
	\begin{aligned}
		k_h &= k_1a_1^*+k_2a_2^*+k_3a_3^*\\
		k_o &= k_1'a_1^{*'}+k_2'a_2^{*'}+k_3'a_3^{*'}
	\end{aligned}
\end{align}
where $(k_1,k_2,k_3)$ and $(k_1',k_2',k_3')$ are the k-points in hexagonal and orthorhombic Brillouin zone, respectively. The quantity $e^{ik.R}$ is invariant to the choice of coordinate system. Making use of the relations (2)-(5)

\begin{equation}
	e^{ik_h.R_h}=e^{ik_o.R_o} \implies k_h.R_h-k_o.R_o=2\pi I
\end{equation}
\begin{equation}
	k_1(l'+m')+2k_2m'+k_3n'-(k_1'l'+k_2'm'+k_3'n')=I
\end{equation}
Finally, the relation between k-points in the hexagonal and orthorhombic reciprocal spaces is obtained as
\begin{align}
	\begin{aligned}	
		k_1' &= k_1 \\
		k_2' &= k_1+2k_2 \\
		k_3' &= k_3
	\end{aligned}
\end{align}
Therefore, the location of $\Gamma$, M and, K points in orthorhombic reciprocal space are (0, 0, 0), (1/2, 1/2, 0), and (1/3, 0, 0), respectively.
\newpage
\section{V\MakeLowercase{ariation of spin splitting with uniaxial strain}}
\begin{figure}[H]
	\includegraphics[width=6in]{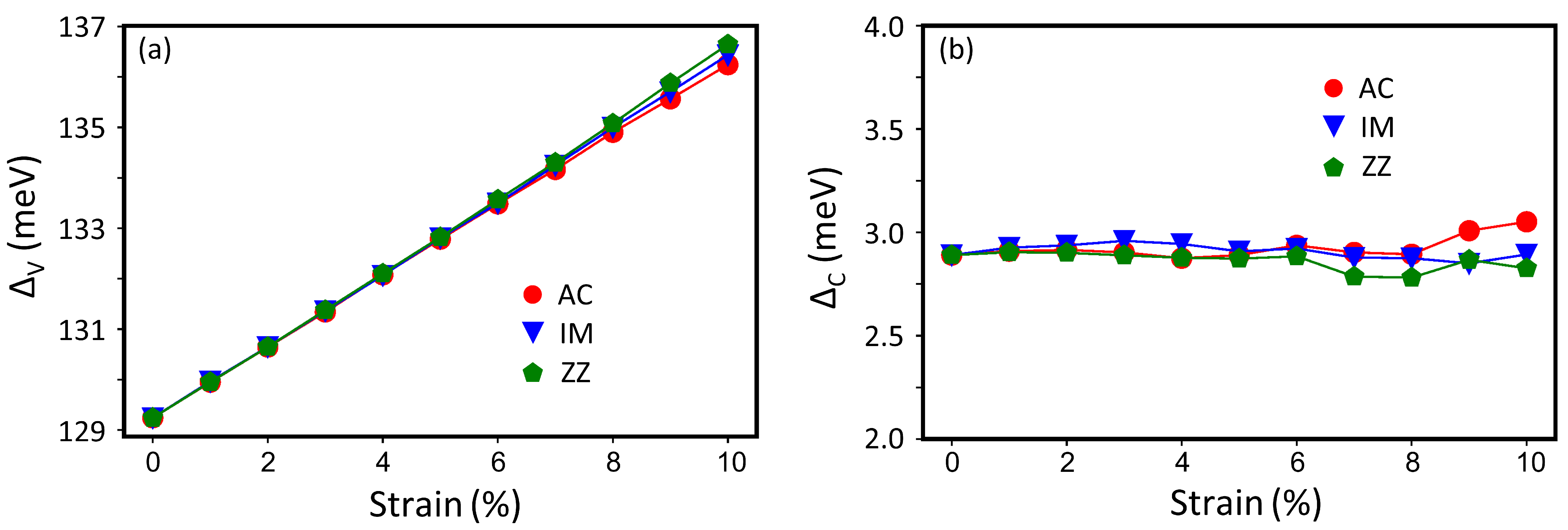}
	\caption{Variation in the spin splitting of (a) VB ($\Delta_{\textrm{V}}$) and (b) CB ($\Delta_{\textrm{C}}$) at the K/K$'$ point driven by spin-orbit coupling with respect to the uniaxial strain along AC, IM, and ZZ directions.}
	\label{figS2}
\end{figure}

The absence of inversion symmetry and strong spin-orbit coupling (SOC) arising in monolayer MoSi$_2$N$_4$ due to the Mo-$4d$ orbitals removes the spin-degeneracy at the arbitrary k-point. Orbital interaction is closely associated with the strain in the lattice structure.  There are several reports on the spin splitting in unstrained and biaxially strained monolayer MoSi$_2$N$_4$~\cite{li2020valley,yang2021valley}. However, the effect of uniaxial strain is still unknown. Given this, the relation between uniaxial strain and spin-valley splitting is also obtained (see Fig.~\ref{figS2}). Without any strain, VB is spin split of the order 130 meV, whereas the CB is nearly spin degenerate. These effects are easily explained by the two band $k.p$ Hamiltonian as derived in Ref.~\cite{PhysRevLett.108.196802} by employing the basis $(|d_{z^2}\rangle, \frac{1}{\sqrt{2}}|d_{x^2-y^2}+i\tau d_{xy}\rangle)\otimes(|\uparrow\rangle, |\downarrow \rangle)$, where $|\uparrow\rangle$ and $|\downarrow \rangle)$ are eigenstates denoting spin-up and spin-down states, respectively, and $\tau=\pm1$ is the valley index. The total Hamiltonian, including SOC, can be expressed as
\begin{equation}
	H= at(\tau k_x \sigma_x+k_y\sigma_y)+\frac{\Delta}{2}\sigma_z-\lambda\tau(\frac{\sigma_z-1}{2})s_z
	\label{H1}
\end{equation}
where $a$ is the lattice constant, $t$ is effective hoping integral, $\Delta$ is the direct gap, and $2\lambda$ is the spin splitting of VB at K point. $\sigma_i$ and $s_i$ are Pauli matrices for the basis functions $(|d_{z^2}\rangle, \frac{1}{\sqrt{2}}|d_{x^2-y^2}+i\tau d_{xy}\rangle) $ and $(|\uparrow\rangle, |\downarrow \rangle)$, respectively. The large spin splitting of the VB is a direct consequence of the complex orbitals arising from the hybridization between Mo-$d_{xy}$ and Mo-$d_{x^2-y^2}$. Spin splitting is absent for CB since only Mo-$d_{z^2}$ contributes. Note that tiny splitting of order 3.0 meV is also observed for CB at the K point since a small portion of N-$p$ orbitals also contributes to CB, which was not included in the Eq.~\ref{H1}. Therefore, a higher number of bands are required in $k.p$ Hamiltonian to explain the small splitting of CB~\cite{kormanyos2015k}. Uniaxial strain modifies the contribution between different orbitals and thus affects the constants ($a, t, \Delta, \lambda$) in the Hamiltonian given by Eq.~\ref{H1}. Therefore, uniaxial strain strongly influences spin-valley splitting. Under the uniaxial tensile strain of range 0-10\%, the VB splitting at K point increases linearly from 129 to 136 meV, whereas VB splitting is nearly independent of the uniaxial strain.
\newpage
\section{E\MakeLowercase{nergy contours in the full} BZ}
\begin{figure}[h]
	\includegraphics[width=4.5in]{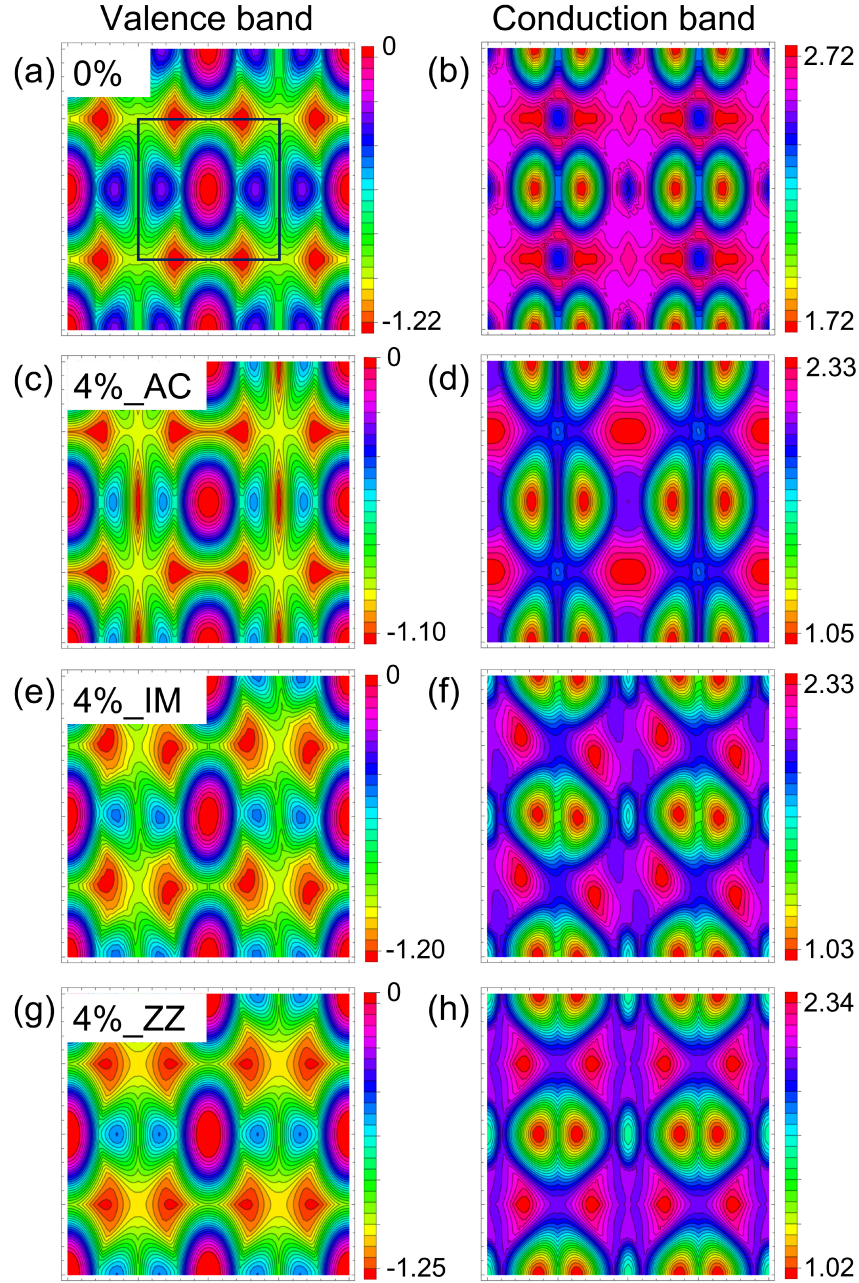}
	\caption{Color map of the (a) VB and (b) CB energies without strain. Similarly, (c)-(h) are the counterparts of (a)-(b) with the uniaxial strain of 4\% along AC, IM, and ZZ directions. Zero energy corresponds to the VBM. The black box in (a) shows the first Brillouin zone. The region considered to plot is the area enclosed by (-$\frac{2\pi}{a_1'}$,-$\frac{2\pi}{a_2'}$) to ($\frac{2\pi}{a_1'}$,$\frac{2\pi}{a_2'}$).}
\end{figure}

\section*{References}
\bibliography{siref}